\documentclass[conference,a4paper]{IEEEtran}
\IEEEoverridecommandlockouts
\usepackage{cite}
\usepackage{amsmath,amssymb,amsfonts}
\usepackage{bm}
\usepackage{algorithmic}
\usepackage{graphicx}
\usepackage{textcomp}
\usepackage[absolute]{textpos}
\usepackage{psfrag}
\usepackage{xcolor}
\usepackage{booktabs}
\usepackage{tikz}
\usetikzlibrary{positioning}
\def\BibTeX{{\rm B\kern-.05em{\sc i\kern-.025em b}\kern-.08em
    T\kern-.1667em\lower.7ex\hbox{E}\kern-.125emX}}

\newcommand{\copyrightstatement}{
    \begin{textblock}{15}(0.5,0.1)    % tweak here: {box width} leftposition, vertPosition)
         \noindent
         \centering
         \textblockcolour{white}
         \footnotesize
         \copyright 2023 IEEE. Personal use of this material is permitted. Permission from IEEE must be obtained for all other uses, in any current or future media, including reprinting/republishing this material for advertising or promotional purposes, creating new collective works, for resale or redistribution to servers or lists, or reuse of any copyrighted component of this work in other works. 
    \end{textblock}}

\begin{document}

\title{Power Reduction Opportunities on End-User Devices in Quality-Steady Video Streaming}

\copyrightstatement

\author{
\IEEEauthorblockN{Christian Herglotz\IEEEauthorrefmark{1}, Werner Robitza\IEEEauthorrefmark{2}\IEEEauthorrefmark{3}, Alexander Raake\IEEEauthorrefmark{3}, Tobias Hossfeld\IEEEauthorrefmark{4}, and Andr\'e Kaup\IEEEauthorrefmark{1}}
\IEEEauthorblockA{\IEEEauthorrefmark{1}Friedrich-Alexander-Universität, Erlangen, Germany --
Email: \emph{firstname.lastname}@fau.de}
\IEEEauthorblockA{\IEEEauthorrefmark{2}AVEQ GmbH, Vienna, Austria --
Email: werner.robitza@aveq.info}
\IEEEauthorblockA{\IEEEauthorrefmark{3}Audiovisual Technology Group, TU Ilmenau, Germany --
Email: \emph{firstname.lastname}@tu-ilmenau.de}
\IEEEauthorblockA{\IEEEauthorrefmark{4}Chair of Communication
Networks, University of W\"urzburg, Germany  --
Email: tobias.hossfeld@uni-wuerzburg.de}
}

%\author{\IEEEauthorblockN{Anonymous}
%\and
%\IEEEauthorblockN{Anonymous}
%\and
%\IEEEauthorblockN{Anonymous}
%\and
%\IEEEauthorblockN{Anonymous}

\IEEEoverridecommandlockouts
\IEEEpubid{\makebox[\columnwidth]{979-8-3503-1173-0/23/\$31.00 \copyright 2023 IEEE \hfill} \hspace{\columnsep}\makebox[\columnwidth]{ }}
\maketitle

\begin{abstract}
This paper uses a crowdsourced dataset of online video streaming sessions to investigate opportunities to reduce the power consumption while considering QoE. For this, we base our work on prior studies which model both the end-user's QoE and the end-user device's power consumption with the help of high-level video features such as the bitrate, the frame rate, and the resolution. On top of existing research, which focused on reducing the power consumption at the same QoE optimizing video parameters, we investigate potential power savings by other means such as using a different playback device, a different codec, or a predefined maximum quality level. We find that based on the power consumption of the streaming sessions from the crowdsourcing dataset, devices could save more than 55\% of power if all participants adhere to low-power settings.  
\end{abstract}

\begin{tikzpicture}[overlay, remember picture]
\path (current page.north) node (anchor) {};
\node [below=of anchor] {%
2023 15th International Conference on Quality of Multimedia Experience (QoMEX)};
\end{tikzpicture}

\section{Introduction}
In the past years, video communication technology entered the lives of people worldwide. Various online video services such as streaming, social media, or teleconferencing are used every day. Global energy consumption related to this technology has reached a significant extent \cite{ShiftFull19,CarbonTrust21}, such that efforts reducing the energy consumption and thus, the carbon impact, are important for our global future. 

%In this direction, a large variety of 
Research has targeted different components in the video communication chain \cite{Afzal23} contributing to energy consumption, including efficient network topologies \cite{Alsharif2019}, transmission standards \cite{Benmoussa16}, encoding bitrate ladders \cite{Katsenou21}, and power-efficient hardware \cite{Gomes22,Farhat22}. On the software side, research has investigated the possibility to reduce processing power by a lower complexity of encoding \cite{Li11,Penny2016} and decoding \cite{Mallikarachchi20}. Finally, the impact of changing high-level video parameters such as bitrate, frame rate, and resolution on power consumption has been evaluated \cite{Herglotz20}. 

In recent work \cite{Herglotz22a}, a crowdsourced dataset from \cite{Robitza20} was used to evaluate two important aspects of streaming videos on YouTube: The quality of experience (QoE) in terms of the subjective quality (mean opinion score, MOS) and the power consumption while streaming. To obtain the two metrics, two models were used. For the QoE, the ITU-T Rec. P.1203 QoE model was employed \cite{Robitza2018,Raake2017}. For the power consumption, a dedicated model was developed using a set of measurements with two different codecs and two different devices. Both models rely on high-level video parameters, i.e., bitrate, resolution, and frame rate. The study revealed that in the crowdsourcing dataset, many streaming sessions used video parameters that are sub-optimal when considering both the power consumption and the QoE. Consequently, it was found that when providing the same QoE, significant power savings of up to $7\%$ could be reached \cite{Herglotz22a}. 

\begin{figure}[tb!]
\centering
\includegraphics[width=.45\textwidth]{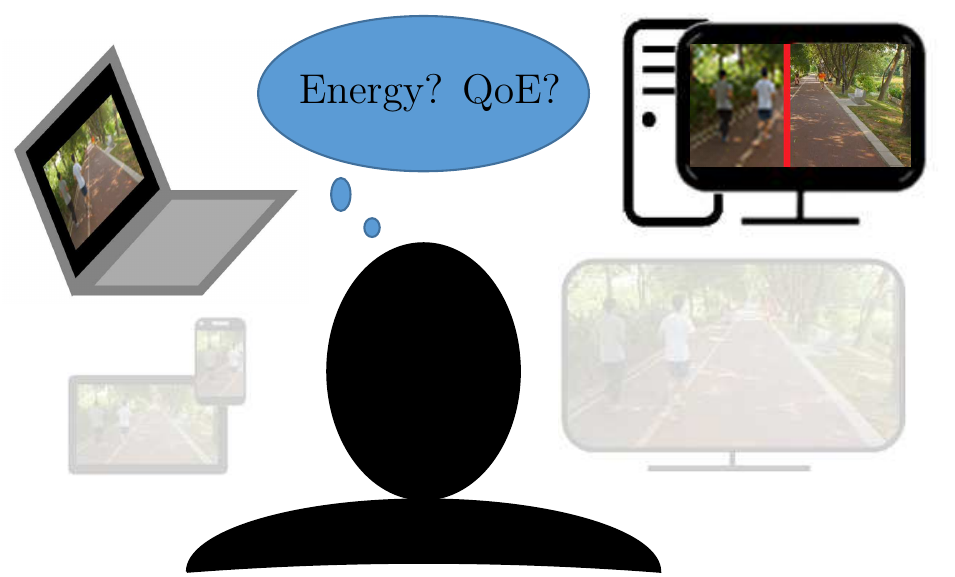}
\vspace{-0.3cm}
\caption{User choices for video quality and device. }
\label{fig:choice}
\vspace{-.5cm}
\end{figure}

In this paper, we used the dataset from \cite{Herglotz22a} to investigate further power reduction opportunities and calculate their impact on the total power consumption. In particular, we assess the influence of the choice of video codec, the playback device, and the use of a maximum QoE as suggested by a sufficiency criterion \cite{Darby18} (see Fig.~\ref{fig:choice}).  Finally, we combine all the approaches and show that up to $55\%$ of power could be saved on end-user devices when participants choose power-efficient video streaming setups. 

%This paper is structured as follows. First, Section~\ref{sec:setup} reviews the quality model, the power model, and the crowdsourced dataset from \cite{Herglotz22a}. Afterwards, in Section~\ref{sec:opp}, we investigate different power-saving strategies under different QoE-constraints. Section~\ref{sec:concl} concludes this paper. 

\section{Setup}
\label{sec:setup}
%\begin{itemize}
%\item  ITU-T P.1203
%\item Power Model (my part)
%\item Crowdsourcing-Dataset
%\item Example result, a scatter plot from \cite{Herglotz22a}
%\end{itemize}

As suggested in~\cite{Herglotz22a}, we calculate QoE scores using the ITU-T Rec. P.1203 model, which estimates the overall streaming QoE based on factors such as loading delay, stalling, video resolution, codec, bitrate, and frame rate. The model considers the overall QoE for sessions of up to 5\,min -- consequently, longer sessions were truncated to provide comparable results. The output is a MOS in the range of 1--5, where 1 is \emph{bad} and 5 is \emph{excellent}.

The power consumption is modeled using the equation
\begin{equation}
    \hat P = \boldsymbol{v}\cdot \boldsymbol{x}^\mathrm{T}, 
\label{eq:powerModel}
\end{equation}
where $\hat P$ is the estimated mean power of a streaming session, $\boldsymbol{v}$ a row vector of high-level video parameters similar to the QoE model (frame rate, pixels per second, bitrate, online/offline scenario, codec), and $\boldsymbol{x}^\mathrm{T}$ a transposed row vector of trained parameters of the same size as $\boldsymbol{v}$. For each device, $\boldsymbol{x}$ was trained separately. We use the trained values from \cite{Herglotz22a}, where mean relative estimation errors below $2\%$ were reported. 

Both QoE and power consumption scores were calculated for the dataset from~\cite{Robitza20}, which comprises 477,000 desktop-based streaming sessions from YouTube in Germany. Note that in this study, we only consider laptops and desktop PCs because the dataset does not provide data for other devices such as TVs or smartphones. 

\section{Power-Saving Opportunities}
\label{sec:opp}
\begin{figure}
\includegraphics[width=.5\textwidth]{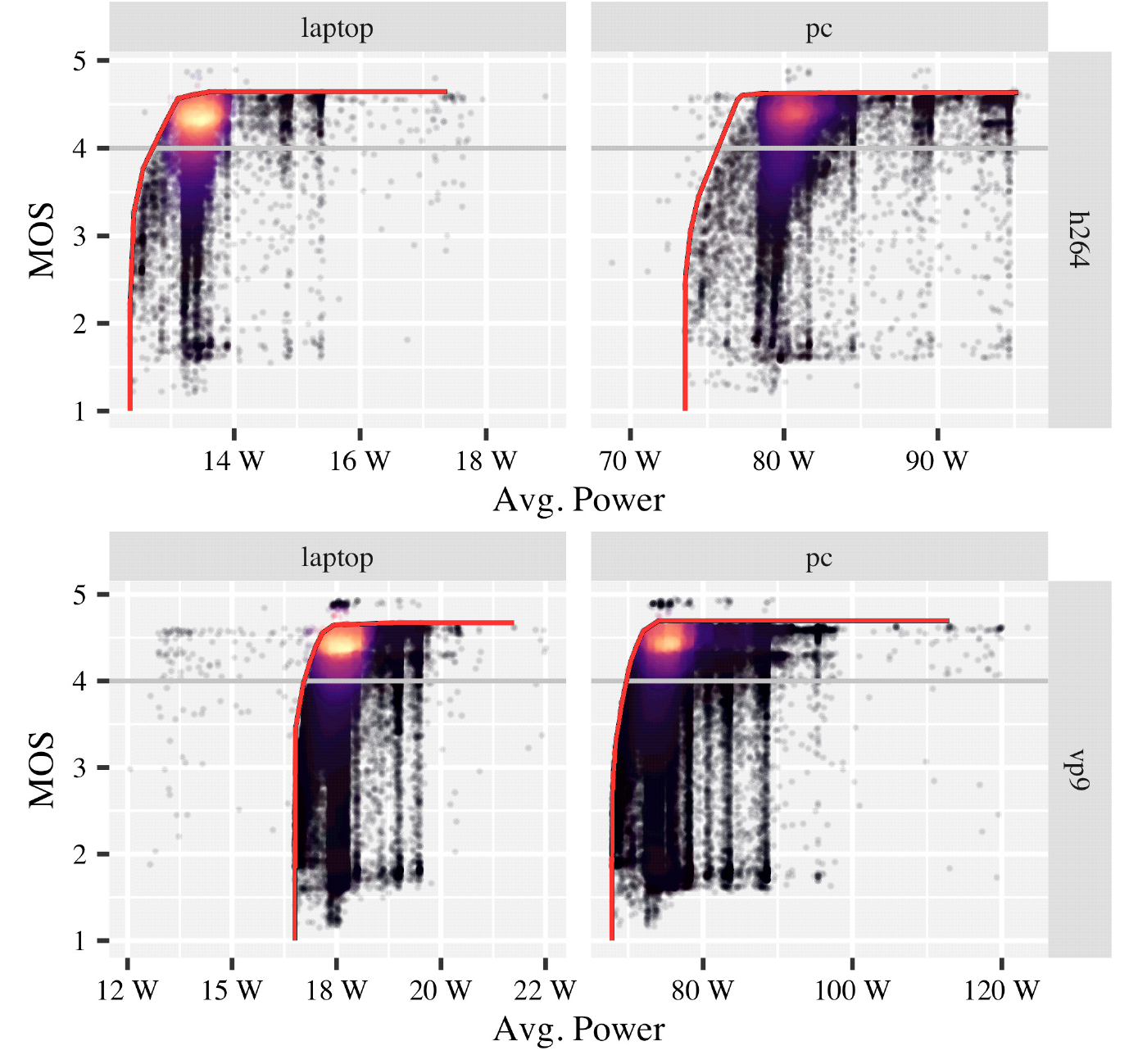}
\vspace{-0.7cm}
\caption{Streaming session MOS values and average device power for two different codecs (VP9, H.264) and devices (laptop, PC) taken from~\cite{Herglotz22a}. Each black point corresponds to a single streaming session. The bright regions correspond to cluster points. The red lines connect Pareto-efficient points.  }
\label{fig:mos_energy}
\end{figure}

In \cite{Herglotz22a}, it was reported that power can be saved by choosing optimal video parameters (parameter optimization). This conclusion can be explained using Fig.~\ref{fig:mos_energy} from \cite{Herglotz22a}. 
In the figure, for each device and codec, the black markers indicate the estimated QoE and estimated power consumption for each streaming session from the crowdsourcing dataset. Then, a Pareto-efficient session was defined by the authors as follows \cite{Herglotz22a}: If there is no other session that is superior in both metrics (MOS and power), it is Pareto-efficient. In red, the Pareto front is drawn, which connects all Pareto-efficient sessions (outliers removed). We can see that many sessions are located below and to the right of the Pareto curve, such that power could be saved at the same MOS. Depending on the device and the video codec, mean power savings between $3\%$ and $7\%$ were reported while the QoE was not affected \cite{Herglotz22a}.% In this paper, we discuss further power saving opportunities that could be obtained on top of these savings.   

\subsection{Optimized Video Parameters}
In \cite{Herglotz22a}, the reason for reported savings were not discussed. Therefore, in this paper, we investigate the relation between the high-level video parameters and the distance to the Pareto front and find that the frame rate is a major factor in power consumption. While many videos are streamed at a high frame rate ($> 24\,$fps), the QoE is often similar at a lower frame rate of $24\,$fps. In high-fps cases, the higher power consumption is then caused by both a higher frame rate and the higher bitrate that is needed to encode higher-fps video with visual quality comparable to that of a lower-fps one.

It should be noted that higher fps videos may indeed offer higher QoE scores, but that effect could not be measured with the existing P.1203.1 video quality estimation module of the P.1203 QoE model, as it does not consider QoE improvements for fps $> 24$. In particular for content like sports or gaming, higher fps values may offer significant QoE benefits at the expense of higher power consumption. Future work may use models from the ITU-T Rec. P.1204 family~\cite{Raake2020}, which provides quality predictions for fps values up to 60.

\subsection{Video Codec Switching}
Results in \cite{Herglotz22a} indicate that the choice of the codec has a significant influence on the power consumption. Therefore, we calculate mean power savings when switching from the less efficient to the more efficient codec. Note that on the laptop, H.264 was the more efficient codec compared to VP9, while on the PC, the contrary was observed. It should be considered that those codec- and device-based savings are dependent on hardware capabilities for decoding video, and therefore the example devices used for modeling in \cite{Herglotz22a} ultimately determine those savings. Future research may cover more devices to provide a more accurate estimation for the power consumption.

To calculate the expected savings, we model the power consumption of all streaming sessions with two parameter sets using Eq.~\eqref{eq:powerModel}: first, with the parameters for the inefficient codec (VP9 for the laptop and H.264 for the PC). Second, we model the power consumption with the efficient codec. Afterwards, we calculate the mean relative difference. The results are listed in Table~\ref{tab:codecSwitch}. 
\begin{table}
\centering
\caption{Relative and absolute end-device power savings when switching from an inefficient to a power-efficient video codec. }
\vspace{-0.3cm}
\label{tab:codecSwitch}
\begin{tabular}{r|rr|rr}
\toprule
 & \multicolumn{2}{c|}{\textbf{Laptop (VP9 to H.264)}} & \multicolumn{2}{c}{\textbf{PC (H.264 to VP9)}}\\
 & Relative & Absolute & Relative & Absolute \\
 \midrule
Savings & $-26.8\%$ & $-4.73\,$W& $-13.4\%$ & $-11.0\,$W\\
\bottomrule
\end{tabular}
\vspace{-0.3cm}
\end{table}
We can find that savings are significantly higher than observed by parameter optimization ($27\%$ for the laptop and $13\%$ for the PC). 

\subsection{Device Switching}
In this subsection, we calculate the power savings when switching from the PC to the low-power laptop using the same video codec. Results are listed in Table~\ref{tab:deviceSwitch}.
\begin{table}
  \centering
  \caption{Relative and absolute end-device power savings when switching from desktop PC to a laptop. }
  \vspace{-0.3cm}
  \label{tab:deviceSwitch}
  \begin{tabular}{r|rr|rr}
  \toprule
   & \multicolumn{2}{c|}{\textbf{H.264}} & \multicolumn{2}{c}{\textbf{VP9}}\\
   & Relative & Absolute & Relative & Absolute \\
   \midrule
  Savings & $-84.2\%$ & $-68.5\,$W& $-77.7\%$ & $-59.3\,$W\\
  \bottomrule
  \end{tabular}
  \vspace{-0.3cm}
\end{table}
As expected, the power savings are high because the base power of the PC is more than $70\,$W and for the laptop less than $20\,$W. The results indicate that for the application of video streaming, handheld devices should be preferred instead of desktop devices, which was also reported in other studies \cite{Hossfeld2023,Afzal23}.

This also raises questions about the everyday use of video streaming services. Lower overall power consumption may be achieved through changing users' habits -- while aiming to retain a good visual quality. For instance, the typical power consumption of large TV screens is significantly greater than for a small tablet device, yet the difference in viewing distance may make a video appear the same size to the user, thus potentially delivering a similar experience. Of course, from a general QoE perspective, those viewing situations may not be directly comparable.

\subsection{Sufficient QoE}
Another possibility to save power is to restrict the QoE (in terms of MOS) to a maximum value. This approach corresponds to the concept of ``energy sufficiency'' \cite{Darby18}, which means that a certain level of quality might be sufficient to satisfy users (the user's ``basic need''), while a significant amount of energy can be saved. Similarly, this concept could be implemented for a green user \cite{Hossfeld2023}, who receives feedback on the power savings caused by the lower visual quality.  The reported power savings then lead to a higher perceived QoE. 

\begin{figure}[t!]
\centering
\includegraphics[width=.45\textwidth]{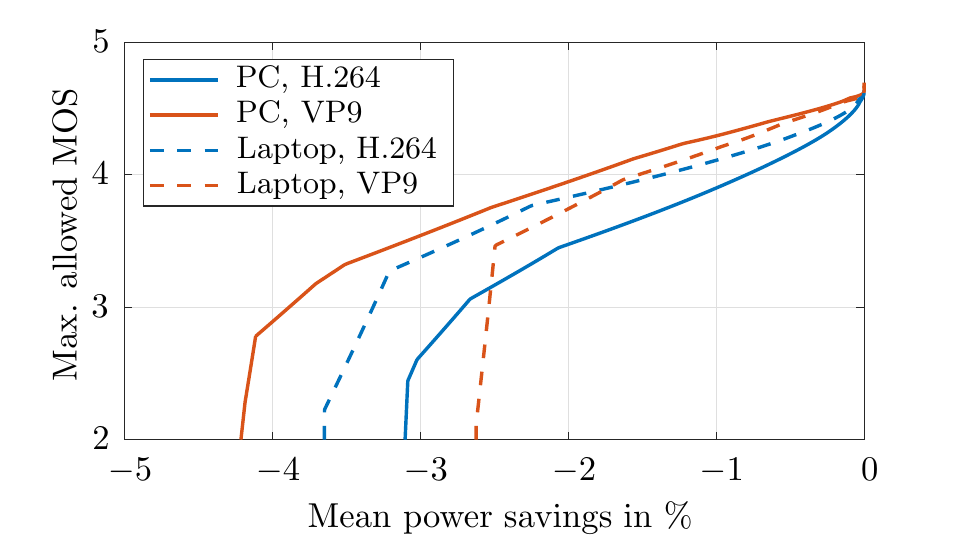}
\vspace{-0.3cm}
\caption{Average savings using energy sufficiency: mean relative power savings in relation to the power otherwise consumed by the end device, depending on the maximum allowed quality. }
\label{fig:sufficientSavings}
\vspace{-0.3cm}
\end{figure}

As such, we calculate relative power savings depending on a maximum MOS and plot the result in Fig.~\ref{fig:sufficientSavings}. 
We can see that by restricting the MOS, power savings by more than $4\%$ can be obtained. Highest savings are achieved at the lowest MOS scores, however, the curves indicate that below a MOS score of $3$, the savings only change marginally. Still, it is unclear which MOS value would correspond to an acceptable quality level as was, e.g., studied in \cite{Li2019AccAnn}. This aspect should be considered for future work. 

Note that lower MOS values in the dataset generally correspond to videos played at lower resolutions and frame rates. It is not intended to artificially lower the MOS by introducing unwanted artifacts like stalling, which would not help consume less power; instead, limitations on visual quality would apply, which could practically be achieved by adding constraints on the bandwidth.

\subsection{Joint Savings}
Finally, we test the potential savings that can be achieved when using the overall optimized setup. This means that for all streaming events in our test set, we choose the optimal streaming configuration (Laptop, H.264) with Pareto-efficient settings (parameter optimization), and compare the resulting power consumption of all streaming sessions with the actual configuration reported in the dataset. Furthermore, we report savings when using an additional sufficiency criterion at a maximum MOS of 3 and 4. The resulting mean power savings over all streaming sessions (including optimized sessions) are listed in Table~\ref{tab:savOverall}. 

% \begin{table}
%   \centering
%   \caption{Relative and absolute end-device power savings when switching all sessions to the most energy efficient configuration. }
%   \label{tab:savOverall}
%   \begin{tabular}{r r|r r|r r}
%   \hline
%    \multicolumn{2}{c|}{Unlimited MOS} & \multicolumn{2}{c|}{$\mathrm{MOS}\le 4$}& \multicolumn{2}{c}{$\mathrm{MOS}\le 3$}\\
%    Rel.  & Abs.  & Rel.  & Abs. & Rel.  & Abs. \\
%    \hline
%    $-55.0\%$ & $-35.7\,$W& $-55.8\%$ & $-35.9\,$W & $-56.8\%$ & $-36.2\,$W \\
%   \hline
%   \end{tabular}
%   \vspace{-0.3cm}
% \end{table}

\begin{table}
  \centering
  \caption{Relative and absolute end-device power savings when switching all sessions to the most energy efficient configuration. }
  \label{tab:savOverall}
  \vspace{-0.3cm}
  \begin{tabular}{rr|rr|rr}
  \toprule
  \multicolumn{2}{c|}{\textbf{Unlimited MOS}} & \multicolumn{2}{c|}{\textbf{MOS}\,$\bm{\le}$\,\textbf{4}}& \multicolumn{2}{c}{\textbf{MOS}\,$\bm{\le}$\,\textbf{3}}\\
   %\multicolumn{2}{c}{Unlimited MOS} & \multicolumn{2}{c}{$\mathrm{MOS}\le 4$}& \multicolumn{2}{c}{$\mathrm{MOS}\le 3$}\\
   Relative  & Absolute  & Relative  & Absolute & Relative  & Absolute \\
   \midrule
   $-55.0\%$ & $-35.7\,$W& $-55.8\%$ & $-35.9\,$W & $-56.8\%$ & $-36.2\,$W \\
  \bottomrule
  \end{tabular}
  \vspace{-0.3cm}
\end{table}

We find that based on the crowdsourcing dataset, more than half of the power consumption could have been saved by choosing optimal video parameters and an energy-efficient device. The major part of these savings can be attributed to the device, i.e., the laptop. Further optimizations (video codec, video parameters, maximum MOS) have a lower impact. It is noteworthy that the maximum MOS seems to have little impact ($55.0\%$ vs. $55.8\%$ savings for unlimited MOS vs. $\mathrm{MOS}\le4$, respectively), which is less than one percentage point and hence smaller than reported in Fig.~\ref{fig:sufficientSavings}. The reason is that savings are calculated from a greater baseline, i.e. the base power consumption of the PC. Consequently, relative savings with respect to this baseline are lower.

%\section{Carbon Impact of this Paper}
%The carbon impact of this publication is estimated based on the work and PC time of writing the manuscript and running evaluations. As the two models as well as the crowdsourcing dataset are already existing, we do not take them into account in the calculation. All in all, a desktop PC was running $10\,$h at an estimated average power consumption of $100\,$W, which results in kgCO2. This corresponds to TODO.   

% Wednesday 1h, THursday 5h, Friday 2 h, Tuesday 2h
% Polishing 3 h
% + 2h Werner
% + 3h rest

\section{Conclusion}
\label{sec:concl}
In this paper, we studied various opportunities to reduce the power consumption on the end-user device while streaming videos. We found that in the case of video streaming, based on the data from a crowdsourcing dataset, more than $50\%$ of the power consumption could have been saved if the most energy-efficient video playback configuration had been used. 

In future work, we will investigate further energy reduction opportunities including other end-user devices such as TVs and smartphones as well as device configurations. We will also investigate the QoE dependency on the playback device's properties and the environment it is used in and how to include the QoE of green streaming. Finally, subjective tests to prove these findings could be performed. 

%\section*{Carbon Emission Statement}
%The generation of this paper has led to roughly  $2\,$kWh of energy consumption (roughly $400\,$g of CO$_2$ emissions). Baseline is the working hours of authors using a desktop PC of $100\,$W. The energy to construct datasets and perform power measurements is not included because they were performed for preceding publications \cite{Robitza20,Herglotz22a}. 

\bibliographystyle{IEEEbib}
\bibliography{IEEEabrv,literature}

\end{document}